\documentclass[aps,prl,twocolumn,showpacs,psfig,superscriptaddress,longbibliography]{revtex4-1}
\usepackage{epsfig}
\usepackage{graphics}
\usepackage{amsfonts}
\usepackage{mathrsfs}
\usepackage{amsmath}
\usepackage{color}
\usepackage{natbib}
\usepackage{textcomp}
\usepackage{graphicx}
\usepackage{bm}
\usepackage{amssymb}
\usepackage{ulem}
\usepackage{xspace}
\usepackage{epstopdf}
\usepackage{dcolumn}
\usepackage{longtable}
\usepackage{subfigure}
\usepackage{multirow}
\usepackage{diagbox}
\usepackage{array}
\usepackage{makecell}
\usepackage[colorlinks=true, letterpaper=true, pdfstartview=FitV, linkcolor=blue, citecolor=blue, urlcolor=blue]{hyperref}
\usepackage{float}

\makeatletter

\newcommand{\Rmnum}[1]{\expandafter\@slowromancap\romannumeral #1@}
\makeatother
\begin{document}

\title{Interlayer Charge-density-wave Vector Phase Induced Structural Chirality}
 
 \author{Sen Shao  }
 \email{These authors contributed equally to this work.}
\affiliation{ Division of Physics and Applied Physics, School of Physical and Mathematical Sciences, Nanyang Technological University, 21 Nanyang Link, 637371, Singapore }

\author{ Wei-Chi Chiu }
\email{These authors contributed equally to this work.}
\affiliation{ Department of Physics, Northeastern University, Boston, MA 02115, USA}
\affiliation{ Quantum Materials and Sensing Institute, Northeastern University, Burlington, MA 01803, USA}

\author{ Tao Hou}
\affiliation{ Division of Physics and Applied Physics, School of Physical and Mathematical Sciences, Nanyang Technological University, 21 Nanyang Link, 637371, Singapore }

\author{ Naizhou Wang}
\affiliation{ Division of Physics and Applied Physics, School of Physical and Mathematical Sciences, Nanyang Technological University, 21 Nanyang Link, 637371, Singapore }

\author{ Ilya Belopolski}
\affiliation{RIKEN Center for Emergent Matter Science (CEMS), Wako, Saitama 351-0198, Japan}

\author{ Yilin Zhao}
\affiliation{ Division of Physics and Applied Physics, School of Physical and Mathematical Sciences, Nanyang Technological University, 21 Nanyang Link, 637371, Singapore }

\author{ Jinyang Ni}
\affiliation{ Division of Physics and Applied Physics, School of Physical and Mathematical Sciences, Nanyang Technological University, 21 Nanyang Link, 637371, Singapore }

\author{ Qi Zhang}
\affiliation{ Laboratory for Topological Quantum Matter and Advanced Spectroscopy (B7), Department of Physics, Princeton University, Princeton, New Jersey 08544, USA}

\author{ Yongkai Li}
\affiliation{ Centre for Quantum Physics, Key Laboratory of Advanced Optoelectronic, Quantum Architecture and Measurement (MOE), School of Physics, Beijing Institute of Technology, Beijing 100081, China}
\affiliation{ Beijing Key Lab of Nanophotonics and Ultrafine Optoelectronic Systems, Beijing Institute of Technology, Beijing 100081, China}
\affiliation{ Material Science Center, Yangtze Delta Region Academy of Beijing Institute of Technology, Jiaxing 314011, China}

\author{ Jinjin Liu}
\affiliation{ Centre for Quantum Physics, Key Laboratory of Advanced Optoelectronic, Quantum Architecture and Measurement (MOE), School of Physics, Beijing Institute of Technology, Beijing 100081, China}
\affiliation{ Beijing Key Lab of Nanophotonics and Ultrafine Optoelectronic Systems, Beijing Institute of Technology, Beijing 100081, China}
\affiliation{ Material Science Center, Yangtze Delta Region Academy of Beijing Institute of Technology, Jiaxing 314011, China}

\author{ Mohammad Yahyavi}
\affiliation{ Division of Physics and Applied Physics, School of Physical and Mathematical Sciences, Nanyang Technological University, 21 Nanyang Link, 637371, Singapore }

\author{ Yuanjun Jin}
\affiliation{ Division of Physics and Applied Physics, School of Physical and Mathematical Sciences, Nanyang Technological University, 21 Nanyang Link, 637371, Singapore }

\author{ Qiange Feng}
\affiliation{ Division of Physics and Applied Physics, School of Physical and Mathematical Sciences, Nanyang Technological University, 21 Nanyang Link, 637371, Singapore }
\affiliation{ Department of Physics, Southern University of Science and Technology, Shenzhen, Guangdong 518055, China}

\author{ Peiyuan Cui}
\affiliation{ Division of Physics and Applied Physics, School of Physical and Mathematical Sciences, Nanyang Technological University, 21 Nanyang Link, 637371, Singapore }

\author{ Cheng-Long Zhang}
\affiliation{ Beijing National Laboratory for Condensed Matter Physics, Institute of Physics, Chinese Academy of Sciences, Beijing 100190, China}

\author{ Yugui Yao}
\affiliation{ Centre for Quantum Physics, Key Laboratory of Advanced Optoelectronic, Quantum Architecture and Measurement (MOE), School of Physics, Beijing Institute of Technology, Beijing 100081, China}
\affiliation{ Beijing Key Lab of Nanophotonics and Ultrafine Optoelectronic Systems, Beijing Institute of Technology, Beijing 100081, China}

\author{ Zhiwei Wang}
\affiliation{ Centre for Quantum Physics, Key Laboratory of Advanced Optoelectronic, Quantum Architecture and Measurement (MOE), School of Physics, Beijing Institute of Technology, Beijing 100081, China}
\affiliation{ Beijing Key Lab of Nanophotonics and Ultrafine Optoelectronic Systems, Beijing Institute of Technology, Beijing 100081, China}
\affiliation{ Material Science Center, Yangtze Delta Region Academy of Beijing Institute of Technology, Jiaxing 314011, China}

\author{ Jia-Xin Yin}
\affiliation{ Department of Physics, Southern University of Science and Technology, Shenzhen, Guangdong 518055, China} 

\author{ Su-Yang Xu}
\affiliation{ Department of Chemistry and Chemical Biology, Harvard University, Cambridge, MA, USA}

\author{ Qiong Ma}
\affiliation{ Department of Physics, Boston College, Chestnut Hill, MA, USA}
\affiliation{ CIFAR Azrieli Global Scholars program, CIFAR, Toronto, Canada}

\author{ Wei-bo Gao}
\affiliation{ Division of Physics and Applied Physics, School of Physical and Mathematical Sciences, Nanyang Technological University, 21 Nanyang Link, 637371, Singapore }

\author{ Md Shafayat Hossain}
\affiliation{ Department of Materials Science and Engineering, University of California, Los Angeles, California, 90095 USA}

\author{ Arun Bansil }
\email{ar.bansil@northeastern.edu}
\affiliation{ Department of Physics, Northeastern University, Boston, MA 02115, USA}
\affiliation{ Quantum Materials and Sensing Institute, Northeastern University, Burlington, MA 01803, USA}

\author{ Guoqing Chang }
\email{guoqing.chang@ntu.edu.sg}
\affiliation{ Division of Physics and Applied Physics, School of Physical and Mathematical Sciences, Nanyang Technological University, 21 Nanyang Link, 637371, Singapore }

\date{\today}


\begin{abstract}
 Chiral charge density waves (CDWs) have attracted intense interest due to their exotic quantum properties, yet the microscopic origin of structural chirality emerging from correlated charge order remains elusive.  Here, we reveal that the interlayer phases of CDW wave vectors, an overlooked degree of freedom, play a crucial role in driving chiral structural displacements in layered CDW materials.  By explicitly incorporating the interlayer phases in first-principles calculations, we successfully obtained the chiral structure of the CDW phases of AV$_3$Sb$_5$ (A= K, Rb, and Cs) and 1T-TiSe$_2$. The electronic and optical properties of the predicted chiral structures are consistent with experimental measurements of these materials in their CDW phases. We further predict that 1T-NbSe$_2$ is a promising material candidate for realizing chiral CDW order.  Beyond materials prediction, our theory reveals that the chiral CDW can be manipulated by electron filling. Our study opens new avenues for discovering, designing, and engineering chiral CDW materials.
\end{abstract}

\maketitle

Chirality, a fundamental geometrical property, plays a pivotal role in understanding a wide range of exotic phenomena in condensed matter systems \cite{PhysRev.104.254, PhysRev.105.1413, rikken2000enantioselective, pendry2004chiral, tang2010optical, skyrmion1, skyrmion2, Nonlocal, Mn3Sn, chang2018topological, hasan2021weyl, yang2021chiral, wang2023chiral, chiralbinhai,gooth2019axionic,li2024signatures,guo2024distinct,szasz2020chiral}. A notable example is a chiral charge density wave (CDW), where chirality arises from a correlated charge order \cite{ishioka2010chiral,van2011chirality,zenker2013chiral, xu2020spontaneous, liu2023electrical, zhao2023spectroscopic, jiang2021unconventional, yang2022visualization, guo2022switchable, vaskivskyi2016fast, zong2018ultrafast, wang2021electronic}. In 1T-TiSe$_{2}$/1T-TaS$_{2}$, fast electrical and optical control and manipulation of chirality have been realized in the chiral CDW phases \cite{vaskivskyi2016fast, zong2018ultrafast, xu2020spontaneous, yang2022visualization,liu2023electrical, zhao2023spectroscopic}. Recent studies of AV$_3$Sb$_5$ (A = K, Rb, Cs) have further demonstrated that chiral CDWs can support unconventional properties in transport, optics, and superconductivity \cite{yang2020giant,jiang2021unconventional, guo2022switchable,wang2021electronic}. Despite significant research interest, the mechanism behind chirality arising from correlated charge orders remains elusive \cite{ishioka2010chiral,van2011chirality,zenker2013chiral, Elmers2025Chirality,tan2021charge, subedi2022hexagonal,Zhang2024Atomis,Kim2024Ori,Kim2024Mic}. 

\begin{figure*}[t]
	\centering
	\includegraphics[scale=0.135]{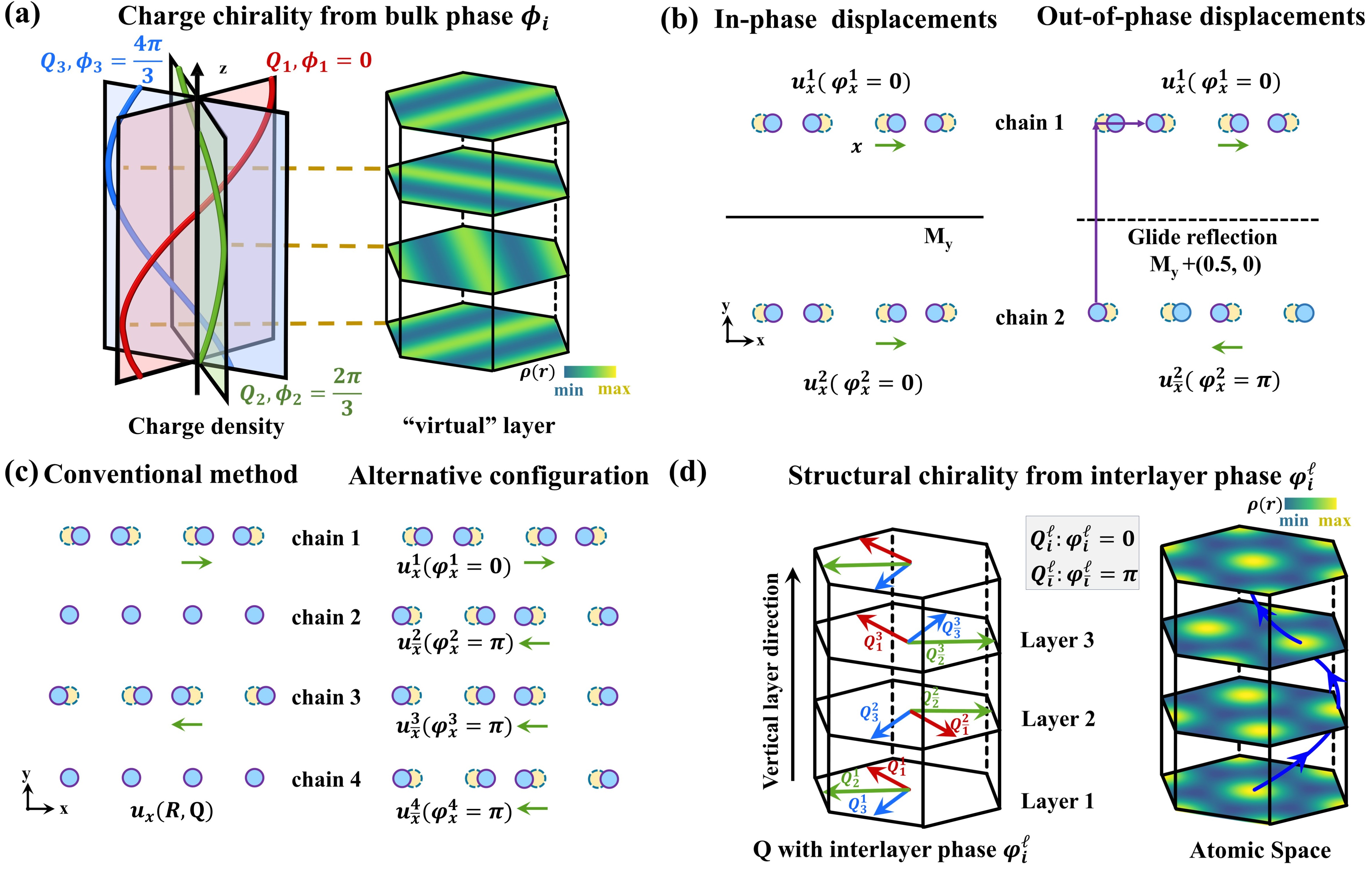}\\
	\caption{(a) Left: Charge density modulation associated with each bulk wave vector $\bm{Q}_i$ associated with different phases $\phi_i$ \cite{ishioka2010chiral}. Right: Schematic of charge distribution in the ``virtual" layer induced by different bulk wave vector phases \cite{ishioka2010chiral}. (b) Schematic of a quasi-1D structural transition induced by in-phase and out-of-phase CDW displacements across chains. Orange and blue circles correspond to atomic positions in the original cell and the CDW cell, respectively. $u^\ell_x$ and $u^\ell_{\overline{x}}$ indicate opposite CDW displacements in real space, induced by $\varphi^\ell_{x}$=0 and $\pi$, respectively. (c) Left: Bulk CDW wave vector $\bm{Q}=\left(\pi / x_0, \pi / 2y_0\right)$ distorts chains 1 ($y= 0$) and 3 ($y= 2y_0$) oppositely while leaving chains 2 ($y= y_0$) and 4 ($y= 3y_0$) undistorted. Right: An alternative configuration enabled by independent interchain phase freedom with $\varphi_x^\ell = \{0, \pi, \pi, \pi \}$. (d) Left: Phase-marked interlayer wave vectors establish spiral patterns. Right: Illustration of the charge distribution at the atomic layer associated with different interlayer wave vector phases $\varphi_i^\ell$. }\label{fig1}
\end{figure*}	

A central challenge lies in clarifying the relationship between chiral CDWs and structural chirality. For instance, while recent experiments on CsV$_3$Sb$_5$ reveal that its chiral CDW phase is structurally chiral \cite{Elmers2025Chirality}, first-principles calculations indicate its energetically stable CDW structures to be achiral \cite{tan2021charge, subedi2022hexagonal,Zhang2024Atomis}. Earlier studies suggested that assigning different phases $\phi_i$ to the three inequivalent bulk CDW wave vectors $\bm{Q}_i$, describing atomic displacements as $\bm{u}_i(\bm{R}) = u_0 \bm{\varepsilon}_i \cos(\bm{Q}_i \cdot \bm{R} + \phi_i)$, can generate a helical charge pattern [Fig. \ref{fig1}(a)]\cite{ishioka2010chiral}. While such phase arrangements can produce charge chirality, they necessarily break crystal symmetries within each layer (e.g., inversion, $C_3$ rotation, and certain mirror planes). In contrast, first-principles calculations on these known chiral CDW materials show that thermodynamically stable configurations emerge only when all symmetry-equivalent triple-$\bm{Q}$ components are incorporated within each layer \cite{tan2021charge, subedi2022hexagonal}.

This discrepancy between theoretical modeling, first-principles calculations, and experimental observations hinders the discovery of new chiral CDW materials. Consequently, only a handful of materials are currently known to host chiral CDWs, and all have been discovered serendipitously rather than through systematic design strategies \cite{ishioka2010chiral, jiang2021unconventional, wang2021electronic}. To advance this rapidly emerging field, it is thus crucial to establish a theoretical framework that can achieve structural chirality that aligns with first-principles calculations and experimental observations
 
In this Letter, we identify a long-overlooked degree of freedom in layered CDW materials, the interlayer phases of the CDW wave vectors, which lie beyond the conventional bulk CDW description. These phases produce CDW displacements that preserve symmetry within each layer but break the overall crystal symmetry, thereby giving rise to atomic chirality. We validate this insight with first-principles calculations on kagome materials AV$_3$Sb$_5$ (A = K, Rb, Cs) to find a chiral structure of 2$\times$2$\times$4 CDW phase. To demonstrate generality, we extend our predictions to the triangular-lattice materials TiSe$_2$ and NbSe$_2$.  Furthermore, a minimal band picture, together with density-functional calculations, shows that electron filling tunes the competition between structure with the same and different interlayer phases configurations, enabling electrical switching of chirality.

Chiral CDW materials found in most recent studies are layered systems with weak interlayer coupling, characterized primarily by in-plane atomic displacements with negligible out-of-plane contributions \cite{tan2021charge,Jin2024phase, lee2019}. With this in mind, we begin with one-dimensional (1D) chains exhibiting Peierls distortions along the x-direction, with weak interchain coupling along the y-direction, where each chain undergoes essentially independent in-chain displacements without out-of-chain contributions.  In this scenario, adjacent chains can exhibit either in-phase displacements, resulting in a 2$\times$1  CDW [Fig. \ref{fig1}(b), left], or out-of-phase displacements in opposite directions, leading to a 2$\times$2  CDW [Fig. \ref{fig1}(b), right].  Here, even though each chain remains structurally equivalent, the out-of-phase CDW displacements can break crystal symmetries [Fig. \ref{fig1}(b)], suggesting that out-of-phase CDW displacements could be the key to understanding structural chirality in chiral CDWs.

The 2$\times$2 CDW phase induced by out-of-phase CDW displacements can be described by a single bulk CDW wave vector $\bm{Q}=\left(\pi / x_0, \pi / y_0\right)$, where $x_0$ and $y_0$ are the lattice constants of the unmodulated cell. For chains located at $y=0$ (chain 1) and $y=y_0$ (chain 2), the atomic displacements can be written as $\bm{u}^{1}(x)=u_0 \cos(Q_x \cdot x+\phi+0)\hat{x}$ and $\bm{u}^{2}(x)=u_0 \cos(Q_x \cdot x+\phi+\pi)\hat{x}$.  Besides the global phase $\phi$ of $\bm{Q}$, each chain exhibits additional phases 0 and $\pi$ that fix the $x$-direction modulations, coming from the $y$-direction CDW wave vector term $Q_y \cdot y_0$.
	
When it comes to more complex CDW displacements, the limitations of conventional bulk CDW wave vector description become apparent. For a 2$\times$4 CDW formed by four 1D chains, the conventional approach prescribes a bulk wave vector $\bm{Q}=\left(\pi / x_0, \pi / 2y_0\right)$. This yields atomic displacements where chains 1 ($y= 0$) and 3 ($y= 2y_0$) are distorted in opposite directions, while chains 2 ($y= y_0$) and 4 ($y= 3y_0$) exhibit no CDW distortions [Fig. \ref{fig1}(c), left]. However, due to the weak interchain coupling, alternative configurations of CDW displacements are allowed for 2$\times$4 CDWs. An example is a case where chains 2, 3, and 4 all exhibit displacements opposite to those of chain 1 [Fig. \ref{fig1}(c), right]. 
Yet, such a configuration cannot be captured in the conventional bulk wave vector approach in which all phase relationships are fixed once the CDW unit cell is determined. 
		
To incorporate this additional degree of freedom stemming from the layered nature of chiral CDW materials, we propose a generalized, layer-dependent form for the charge modulation in three-dimensional layered systems:
\begin{equation}
		\bm{u}_i^\ell (\bm{r^{\parallel}} )=u_0 \bm{\varepsilon}_i  \cos(\bm{q}_i \cdot \bm{r}^{\parallel} + \phi_i + \varphi_i^\ell),  \notag
\end{equation}
where $\ell$ indexes the layer, $\bm{q}_i$ is the associated $i$-th CDW wave vector for the layer, $\bm{r}^{\parallel}$ represents the in-plane atomic position within layers, and $\varphi_i^\ell$ is the additional interlayer phase shift for the $i$-th CDW wave vector in layer $\ell$. 
Our key insight is that different choices of interlayer phases $\{\varphi_i^\ell\}$ introduce different CDW displacements between the layers, thereby breaking bulk inversion symmetry and giving rise to structural chirality.

To efficiently represent configurations with different interlayer phases, we define phase-tagged in-plane wave vectors that incorporate interlayer phase information: $\bm{q}_i$ refers to wave vectors with $\varphi_i^\ell = 0$, while $\bm{q}_{\overline{i}}$ refers to those with $\varphi_i^\ell = \pi$, with $\bm{q}_i$ ($\bm{q}_{\overline{i}}$) pointing in the positive (negative) $i$-direction. Consider the case of a layered CDW material with the following combination of interlayer phases configurations: $\varphi^1=[0,0,0]$, $\varphi^2=[\pi,\pi,0]$, and $\varphi^3=[0,\pi,\pi]$. 
The resulting in-plane, phase-tagged $\bm{q}$-vectors form a chiral pattern along the stacking direction [Fig. 1(d), left], corresponding to a structural chiral configuration in real space [Fig. \ref{fig1}(d), right].  

 \begin{figure}[bp]
	\centering
	\includegraphics[scale=0.125,angle=0]{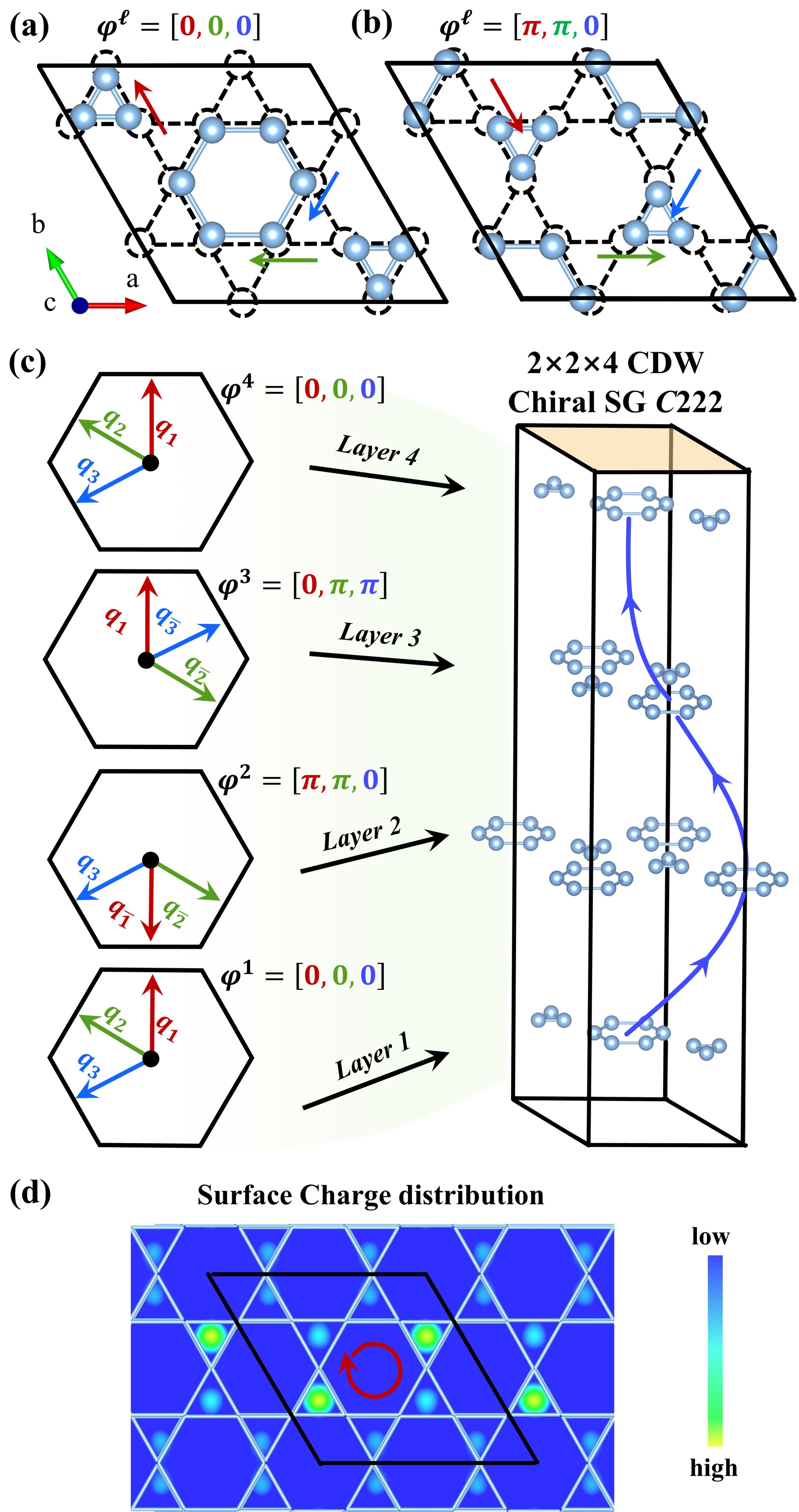}\\
	\caption {(a)-(b) Inverse star-of-David patterns resulting from interlayer phases configurations $\varphi^\ell$= [0, 0, 0] and [$\pi, \pi$, 0], respectively. Black dashed lines represent the ideal kagome lattice. Schematic atomic displacements are denoted by red, blue, and green arrows, respectively.  (c) Incorporation of interlayer wave vector phases in 2$\times$2$\times$4 AV$_{3}$Sb$_{5}$ (A= K, Rb, and Cs) CDWs to induce structural chirality is illustrated. Left: Spiral phase-marked interlayer wave vector. Right: Atomic structure in the chiral space group $\textit{C}$222. (d) Simulated surface charge distribution on the Sb surface.} \label{fig2}
\end{figure}

Next, we apply the preceding approach to the real material CsV$_{3}$Sb$_{5}$ for validation using first-principles calculations. The pristine phase of CsV$_{3}$Sb$_{5}$ lies in a high-symmetry space group $\textit{P}$6/$\textit{mmm}$. Within each layer, there are two possible distorted 2$\times$2 CDW structures: the star-of-David (SoD) and the inverse star-of-David (ISD), induced by three wave vectors $Q_{i}$ (i = 1, 2, 3) \cite{tan2021charge}. Accounting for all possible combinations of the associated interlayer phases configurations [$\varphi_1^\ell$, $\varphi_2^\ell$, $\varphi_3^\ell$] yields four ISD and four SoD CDW patterns for a layer (Figs. S1-S2).
As representative examples, the ISD pattern in Fig. \ref{fig2}(a) corresponds to the interlayer phases configuration [0,0,0], while the pattern in Fig. \ref{fig2}(b), with interlayer phases configuration $[\pi, \pi, 0]$, features CDW displacements that are reversed along directions 1 and 2 relative to the reference, with direction 3 unchanged.
 
By systematically enumerating all inequivalent combinations of interlayer phases configurations and incorporating the corresponding structures into first-principles calculations, we find that the 2$\times$2$\times$2 CDW phase of CsV$_3$Sb$_5$ energetically favors achiral structures in the space group $Fmmm$ [Table S1 in Supplementary Material (SM) \cite{ref-SM}]. This result for the 2$\times$2$\times$2 CDW phase is consistent with previous studies, which show that ISD patterns have lower energy than the SoD patterns \cite{tan2021charge}. Based on this energy preference, in investigating CsV$_{3}$Sb$_{5}$ 2$\times$2$\times$3 and 2$\times$2$\times$4 supercells below, we will focus exclusively on the ISD-type CDW distortions. Our in-depth first-principles calculations encompassing all inequivalent combinations of interlayer phases configurations reveal three thermodynamically as well as dynamically stable chiral ground states: two in the 2$\times$2$\times$3 CDW phase and one in the 2$\times$2$\times$4 phase (Tables S2-S4).

\begin{table}[t] 
	
	\centering
	\caption{Interlayer CDW phases, and their symmetries and enthalpy differences ($\Delta$E) relative to the pristine phases for various layered materials. } 
	\renewcommand{\arraystretch}{1.3}
	\begin{tabular}{ m{1.8cm}<{\centering} m{2cm}<{\centering} m{2cm}<{\centering} m{2cm}<{\centering} }		
		
		\Xhline{1.3pt}
		
		\multirow{2}{*}{Materials}   & {Interlayer} & Symmetry & $\Delta$E     \\
		& Phases & (Cell) & (meV/atom) \\
		
		\hline
		\multirow{2}*{CsV$_{3}$Sb$_{5}$} & $\varphi^1$=[0, 0, 0]& $\textit{F}$$\textit{mmm}$ & \multirow{2}*{-3.7}   \\
		& $\varphi^2$=[$\pi$, $\pi$, 0] &  (2$\times$2$\times$2) &  \\
		
		\hline
		\multirow{4}*{CsV$_{3}$Sb$_{5}$} & $\varphi^1$=[0, 0, 0] & \multirow{3}*{$\textit{C}$222}  & \multirow{4}*{-3.7}   \\
		
		& $\varphi^2$=[$\pi$, $\pi$, 0] &  & \\
		& $\varphi^3$=[0, $\pi$, $\pi$] &  \multirow{2}*{(2$\times$2$\times$4)}  &  \\
		& $\varphi^4$=[0, 0, 0] &  &  \\
		\hline
		\multirow{2}*{1T-TiSe$_{2}$} &  $\varphi^1$=[0, 0, 0]& \multirow{1}*{$\textit{C}$2}  & \multirow{2}*{-1.2}   \\
		& $\varphi^2$=[0, $\pi$, $\pi$] & \multirow{1}*{(2$\times$2$\times$2)} &  \\

		\hline
		\multirow{3}*{1T-NbSe$_{2}$} & $\varphi^1$=[0, 0, 0] & \multirow{2}*{$\textit{P}$3$_{1}$} & \multirow{4}*{-12.2}   \\
		& $\varphi^2$=[$\pi$, 0, 0] & \multirow{2}*{($\sqrt{13}$$\times$$\sqrt{13}$$\times$3)}  &  \\
		& $\varphi^3$=[$\pi$, $\pi$, 0] &    &  \\
		
		\Xhline{1.3pt}
	\end{tabular}\label{tab1}
\end{table}

We note that the 2$\times$2$\times$4 CDW is characterized by the combination of interlayer phases configurations $\varphi^1 = [0, 0, 0]$, $\varphi^2 = [\pi, \pi, 0]$, $\varphi^3 = [0, \pi, \pi]$, and $\varphi^4 = [0, 0, 0]$. Fig. \ref{fig2}(c) shows that the in-plane phase-tagged $\bm{q}$-vectors of this structure rotate along the stacking direction and drive structural chirality in the chiral space group $C$222. We obtained similar results in KV$_{3}$Sb$_{5}$ and RbV$_{3}$Sb$_{5}$ (Tables S6-S11).

Our prediction of 2$\times$2$\times$4 chiral structures is supported by recent X-ray photoelectron diffraction measurements that confirm the presence of chiral 2$\times$2$\times$4 CDW structures in CsV$_{3}$Sb$_{5}$ \cite{Elmers2025Chirality}. Furthermore, our simulated surface charge distributions exhibit spiral patterns [Fig. \ref{fig2}(d)] in good agreement with scanning tunneling microscopy (STM) images \cite{jiang2021unconventional}. Additionally, circular photogalvanic effect measurements indicate that the CDW state of this family is likely to belong to the $C$2 or $D$2 point group \cite{Cheng2025Broken}, consistent with our predicted chiral structure in space group $C$222, which falls within the $D$2 point group.

Our calculations show that the 2$\times$2$\times$2 achiral and 2$\times$2$\times$4  chiral CDW structures are nearly degenerate both in the ideal case (Tables S1,S3) and in crystals with defects (Table S5), see Table \ref{tab1}. This result is consistent with experimental observations, where some experiments exclusively find the 2$\times$2$\times$2 phase \cite{li2021observation}, while others find the 2$\times$2$\times$4 phase \cite{ortiz2021fermi}, and some experiments document coexistence of these two phases \cite{xiao2023coexistence}.

We also applied our framework to consider different choices of interlayer phases $\varphi_i^\ell$ in the transition-metal dichalcogenide (TMD) 1T-TiSe$_2$ (Table \ref{tab1}). Here we obtain a stable chiral structure for 1T-TiSe$_2$ (Table S12), and our predicted structure matches the reported STM image (Fig. S3). \cite{Kim2024Mic}

Beyond predicting chiral structures in known chiral CDW materials, we investigated the triangular-lattice 1T-NbSe$_2$ (\textit{P}$\overline{3}$\textit{m}1), where the 2D chiral CDW has been experimentally observed \cite{song2022atomic}, but the 3D chiral CDW phase remains to be identified. Here we find that 1T-NbSe$_2$ exhibits $\sqrt{13} \times \sqrt{13} \times 3$ chiral CDW phase, whose combination of interlayer phases configurations is: $\varphi^1 = [0, 0, 0]$, $\varphi^2 = [\pi, 0, 0]$, and $\varphi^3 = [\pi, \pi, 0]$. Our thermodynamic and dynamic stability analyses confirm that this chiral structure is stable and energetically favorable. Thus, our calculations predict potential CDW-driven chiral phases in 3D 1T-NbSe$_2$ at low temperatures (Table \ref{tab1}, Table S13).

\begin{figure}[tbp]
	\centering
	\includegraphics[width=0.44\textwidth]{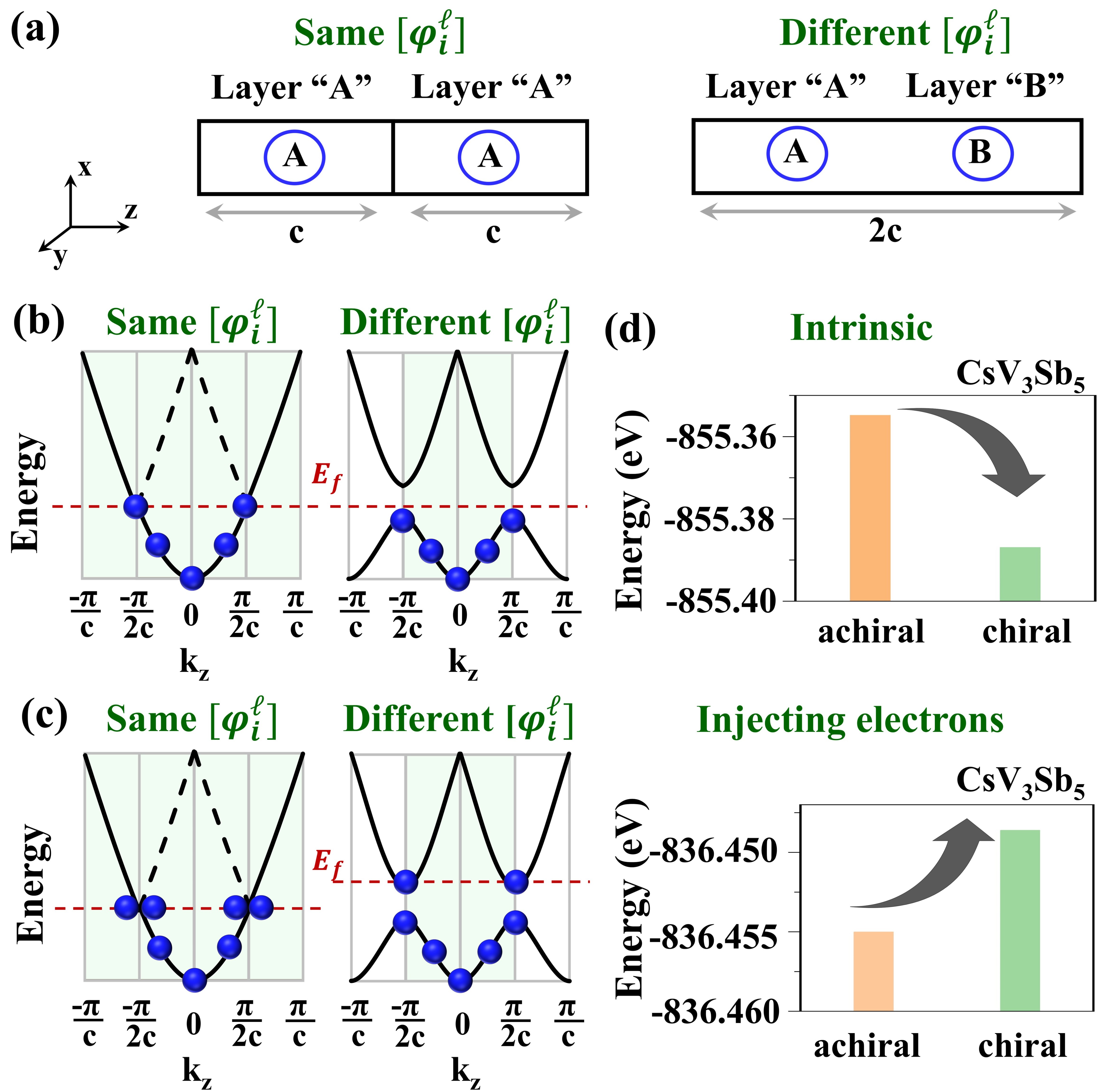}\\
	\caption{(a) Left: Schematic structure with the same interlayer phases configuration ([$\varphi_i^\ell$]) along the z direction, where only layer ``A'' is present. Right: Schematic structure with different [$\varphi_i^\ell$] along the z direction, where layers ``A'' and ``B'' are present. (b) Left: Schematic band for a structure with the same [$\varphi_i^\ell$]. Black solid line shows the single parabolic band along $k_z$ in the original Brillouin zone (BZ), while the black dashed line shows its folded part in the reduced BZ, where $k_z$ = [-$\pi$/2c, $\pi$/2c]. Right: A schematic band for a structure with different [$\varphi_i^\ell$]. Blue spheres denote the electron occupancy of the band. Green shadow zone represents the true BZ of the CDW structure with the same [$\varphi_i^\ell$] (left) and different [$\varphi_i^\ell$] (right). (c) Schematic band structures and the corresponding electron occupancies for structures with the same and different [$\varphi_i^\ell$] in the presence of one additional electron relative to the case of panel (b). (d) Energies for achiral-2$\times$2$\times$1 (same [$\varphi_i^\ell$]) and chiral-2$\times$2$\times$4 (different [$\varphi_i^\ell$]) CsV$_{3}$Sb$_{5}$ CDW structures at the intrinsic Fermi level (top) and with the injection of three additional electrons (bottom). For comparison, we employed the same 2$\times$2$\times$4 cell in the calculations.  } \label{fig3}
\end{figure}

We emphasize that beyond predicting chiral CDW phases in layered materials, our framework also provides a pathway for tuning the chirality through electron filling. The key lies in controlling the relative interlayer phase along the z-axis. When all layers share the same CDW phase (same [$\varphi$$_i^\ell$]), they are translationally equivalent along the z direction [Fig. \ref{fig3}(a), left]. For simplicity, we use a parabolic dispersion to describe the associated band along $k_z$ ([-$\pi$/c, $\pi$/c]) [Fig. \ref{fig3}(b), left]. When a relative CDW phase shift is introduced between the layers (different [$\varphi$$_i^\ell$]), the translational symmetry along z is modified and the layers become inequivalent. As an example, consider the case where the periodicity along the z-axis is doubled, leading to two inequivalent layers [Fig. \ref{fig3}(a), right].
To directly compare the bands of the two cases, we fold the band structure of CDW with the same [$\varphi$$_i^\ell$] into an artificially doubled unit cell, where the two folded bands are degenerate at k$_z$=$\pi$/2c [Fig. \ref{fig3}(b), left and Fig. S4]. By contrast, in the presence of different [$\varphi$$_i^\ell$], the CDW modulation along z couples to the interlayer electronic degrees of freedom and lifts this degeneracy, splitting the two folded bands by pushing one upward and the other downward in energy [Fig. \ref{fig3}(b), right].
This splitting leads to distinct energetic consequences depending on electron filling. For example, when a single electron occupies the originally degenerate folded band point, the splitting lowers the total energy, thereby favoring the structure with different [$\varphi_i^\ell$] [Fig. \ref{fig3}(b)]. In contrast, when two electrons occupy the same band point, the splitting increases the total energy, favoring the structure with the same [$\varphi_i^\ell$] [Fig. \ref{fig3}(c)].

We now apply this idea to the real material CsV$_3$Sb$_5$. At the pristine Fermi level of CsV$_3$Sb$_5$, the chiral CDW structure is energetically favored over the achiral CDW structure [Fig. \ref{fig3}(d), top]. However, upon electron doping, specifically by introducing three additional electrons per unit cell, the energetic hierarchy reverses, and the achiral CDW structure becomes the ground state [Fig. \ref{fig3}(d), bottom].  Thus, the structural chirality of CsV$_{3}$Sb$_{5}$ can be switched off via electrostatic gating or chemical doping. This gate-controllable structural chirality elevates chirality to a controllable order parameter, opening a broad avenue to symmetry-programmable properties and emergent behaviors in chiral quantum matter.

Finally, we comment on the connection between the present interlayer-phase description and the standard crystallographic framework in which any modulated CDW structure can, in principle, be described through a Fourier synthesis involving an appropriate number of higher-order modulation wave vectors (Sec. X in SM). Crystallographic methods treat Fourier coefficients as fitting parameters for post hoc structural characterization. In sharp contrast, our approach restricts interlayer phases to binary values dictated by phonon analysis-reducing the configuration space to a discrete, enumerable set to enable predictive first-principles structure searches.

By uncovering the long-overlooked interlayer phase degree of freedom in charge-density waves, an aspect that has been absent from the conventional bulk $\bm{Q}$-vector descriptions, our study bridges theory and experiment to enable systematic first-principles predictions of structurally chiral CDW materials. Moreover, our introduction of the concept of interlayer-phase-driven symmetry breaking provides a general pathway for designing and engineering new correlated CDW phases in layered materials for hosting emergent quantum phenomena. 


\textit{Acknowledgements} We would like to acknowledge useful discussions with F. Duncan M. Haldane. Work at Nanyang Technological University was supported by the National Research Foundation, Singapore, under its Fellowship Award (NRF-NRFF13-2021-0010), the Singapore Ministry of Education (MOE) Academic Research Fund Tier 3 grant (MOE-MOET32023-0003), and the Agency for Science, Technology and Research (A*STAR) under its Manufacturing, Trade and Connectivity (MTC) Individual Research Grant (IRG) (Grant No.: M23M6c0100).

\end{document}